\documentclass[aps,prl,superscriptaddress,nofootinbib,showkeys,twocolumn,preprintnumbers]{revtex4-1}
\pdfoutput=1
\usepackage[utf8]{inputenc}
\usepackage{uniinput}
\usepackage{graphicx}
\usepackage{color}
\usepackage{amsmath,amssymb,amsthm,mathtools,dsfont,amsfonts}
\usepackage{comment,braket,array,bold-extra,cancel,soul}
\usepackage{enumerate,xargs}
\usepackage{hyperref}
\usepackage[dvipsnames]{xcolor}
\usepackage{epstopdf}
\usepackage{float}
\usepackage{subcaption}
\usepackage{physics}

\newcommand{\be}{\begin{equation}}
\newcommand{\ee}{\end{equation}}

\def\({\left (}
\def\){\right )}

\def\Label#1{\label{#1}%
  \smash{\hbox to0pt{\raise1ex\hbox{\tiny[#1]}\hss}}}
\def\noLabels{\let\Label=\label}
\def\nobbibitem{\let\bbibitem=\bibitem}

\begin{document}
\noLabels
\nobbibitem

\title{Brickwall, Normal Modes and Emerging Thermality}
\author{Souvik \surname{Banerjee}}
\affiliation{\it Institut für Theoretische Physik und Astrophysik, Julius-Maximilians-Universität Würzburg,\\ Am Hubland, 97074 Würzburg, Germany}
\author{Suman \surname{Das}}
\affiliation{\it Theory division, Saha Institute of Nuclear Physics, A CI of Homi Bhabha National Institute, \\
1/AF, Bidhannagar, Kolkata 700064, India.}
\author{Moritz \surname{Dorband}}
\affiliation{\it Institut für Theoretische Physik und Astrophysik, Julius-Maximilians-Universität Würzburg,\\ Am Hubland, 97074 Würzburg, Germany}
\author{Arnab \surname{Kundu}}
\affiliation{\it Theory division, Saha Institute of Nuclear Physics, A CI of Homi Bhabha National Institute, \\
1/AF, Bidhannagar, Kolkata 700064, India.}

\begin{abstract}
\noindent In this article, we demonstrate how black hole quasi-normal modes can emerge from a Dirichlet brickwall model normal modes. We consider a probe scalar field in a BTZ geometry with a Dirichlet brickwall and demonstrate that as the wall approaches the event horizon, the corresponding poles in the retarded correlator become dense and yield an effective branch-cut. The associated discontinuity of the correlator carries the information of the black hole quasi-normal modes. We further demonstrate that a non-vanishing angular momentum non-perturbatively enhances the pole-condensing. We hypothesize that it is also related to quantum chaotic features of the corresponding spectral form factor, which has been observed earlier. Finally, we discuss the underlying algebraic justification of this approximate thermalization in terms of the trace of the algebra.
\end{abstract}

\preprint{XXXXXX}

\maketitle


\section{Introduction \& Discussion}
\noindent One of the earliest achievements of the AdS/CFT correspondence \cite{Maldacena:1997re,Witten:1998qj,Gubser:1998bc} had been to understand black holes in Anti-de Sitter spacetime in terms of distinguishing features of thermal correlators of a non-gravitational conformal field theory living on a codimension one conformal boundary. 

\noindent In this work we study a toy model of a (quantum) black hole spacetime capped off by a ``brick-wall" through an appropriate boundary condition, right before the horizon \cite{tHooft:1984kcu}. This wall is hypothesized to move infinitesimally close to the horizon. This capped-off geometry corresponds to a pure state in the dual CFT, which, qualitatively, resembles a smooth fuzzball microstate \cite{Lunin:2001jy, Kanitscheider:2007wq, Bena:2015bea} in the supergravity approximation \cite{Das:2022evy, Das:2023xjr}.\footnote{It is non-trivial to define a ``smooth geometry" in the highly quantum regime, since a geometric description may not exist in that regime. 
 Nonetheless, it may still be true that an ``effective geometric" description exists that somehow captures quantum gravitational features. These are speculative statements at this point.} In this work we investigate how an effective thermal description emerges in this set-up, in a sense that a low-energy asymptotic observer would never be able to distinguish it from an actual black hole. This exemplifies the perturbative dynamics of thermalization in a typical pure state which lies at the heart of the black hole information paradox \cite{ Susskind:1993if, Kiem:1995iy, Mathur:2005zp, Almheiri:2012rt, Almheiri:2013hfa, Marolf:2013dba, Papadodimas:2012aq, Papadodimas:2013jku, Papadodimas:2013wnh}. 
 
In this brickwall model, we compute the holographic boundary Green's function of an operator dual to a bulk probe scalar field by imposing a Dirichlet boundary condition at the radial cutoff near the event horizon. We demonstrate that when the wall is placed at a distance $\epsilon$ away from the horizon, an asymptotic observer will observe an effective thermality in the limit $\epsilon \to 0$ and will also be able to read off the associated quasi-normal modes (QNM) from the normal modes. Here, $\epsilon$ is a dimensionless number measured in a suitable unit, {\it e.g.}~the Planck length/string length. This cutoff translates to a divergent IR time-scale: $\delta=\log \epsilon$. The QNMs emerge from the inability to access this IR-divergent time scale. In other words, given an arbitrarily long measurement time, one can always find an $\epsilon$ such that the information of the normal modes will repackage itself into a set of QNM that coincides with the black hole QNM. Thus, our observations provide stronger and complementary evidence that a brickwall model can indeed capture the thermal features of a black hole as an effective description. 
{\it In this way, our analyses provide strong evidence that resolving the reflecting Dirichlet wall from a classical event horizon with ingoing boundary conditions is extremely difficult for a low-energy asymptotic observer, as the wall approaches infinitesimally close to the horizon.}

The time scale mentioned above is, largely, kinematical and holds for a linear spectrum in a two-dimensional black hole geometry \cite{Soni:2023fke}. We will show that the IR-divergent time-scale is even non-perturbatively further from $\delta$, once we include non-vanishing angular momenta along the compact direction. Curiously, this non-perturbative separation appears related to the appearance of quantum chaos, as discussed in \cite{Das:2022evy, Das:2023ulz, Das:2023yfj, Das:2023xjr, Krishnan:2023jqn}.  

{\it The main achievement in this work is to provide a novel holographic description of the emergence of thermality in a perturbative setting.} While this is mostly accepted in the literature as a reasonable conjecture, approximation of a typical pure state as a thermal state is demonstrated explicitly in a controllable setting in this work.
The probe scalar and its subsequent quantization is simply a semi-classical perturbative analysis, as we move the Dirichlet wall infinitesimally close to the horizon. The dependence of thermality on $\epsilon$ is similar to \cite{Giusto:2023awo}, but non-perturbatively enhanced. The thermalization is manifest through accumulations of poles of the holographic retarded Green's function on the real axis. As $\epsilon \to 0$, the poles condense and effectively form a branch-cut in the continuum limit.  To resolve the branch-cut from a collection of dense poles one needs an incredibly large time-domain compared to the proper distance of the Dirichlet wall from the event horizon. {\it In this work, we also argue how this scale of resolution determines the type of von Neumann algebra of local operators of the boundary field theory}. This completes an algebraic understanding of a thermal approximation of a typical pure state in a large system.

The rest of the paper is organized as follows. We start with the standard computation of holographic Green's function \cite{Son:2002sd}, but with the Dirichlet condition on the wall. We further discuss the analytic properties of Green's function and demonstrate the effective thermality. Finally, we conclude with a discussion on how this approximate thermalization can be realized in terms of the classification of von Neumann algebra. In particular, we demonstrate how the aforementioned limiting procedure captures an approximate transition from type I to type III von Neumann algebra.


\section{Scalar field in the brickwall}

Let us begin with the non-rotating BTZ \cite{Banados:1992wn} geometry
\begin{equation}\label{metric}
    ds^2=-(r^2-r_{\rm H}^2)dt^2+\frac{dr^2}{(r^2-r_{\rm H}^2)}+r^2 d\psi^2 \,,
\end{equation}
where $r= r_{\rm H}$ is the position of the horizon.
Consider a probe scalar field of mass $\mu$ in this background that satisfies the Klein-Gordon equation
\begin{equation}\label{eom1}
    \Box \Phi\equiv \frac{1}{\sqrt{|g|}}\partial_{\nu}\left(\sqrt{|g|}\partial^{\nu}\Phi\right)=\mu^2 \Phi.
\end{equation}
One can solve \eqref{eom1} analytically with the ansatz, $ \Phi=\sum_{\omega, m}e^{-i\omega t}e^{i m\psi}{\tilde\phi}(r)$. 
In terms of the redefined radial coordinate $z= 1- {r_{\rm H}^2}/{r^2}$ with $z \in [0,1]$, we get two linearly independent solutions in terms of hypergeometric functions. The general solution is a linear combination:
\begin{align}\label{sol1}
    {\tilde\phi}\left(r(z)\right) & \equiv \phi\left(z\right)  =(1-z)^{\beta}\left[ C_1\, z^{-i \alpha} {}_2F_1(a, b; c; z) \right. \nonumber \\
    & \left. + \,  C_2\, z^{i \alpha} \,{}_2F_1(1+a-c, 1+b-c; 2-c; z)\right]
\end{align}
where, 
\begin{align}\label{ab}
  &  a=\beta-\frac{i}{2r_{\rm H}}(\omega + m), \ b=\beta-\frac{i}{2r_{\rm H}}(\omega - m), \ c=1-2i\alpha, \nonumber \\
  & \text{with}  \ \ \
   \alpha=\frac{\omega}{2r_{\rm H}}, \ \ \beta=\frac{1}{2}(1-\sqrt{1+\mu^2}).
\end{align}

We now impose a Dirichlet wall at a radial position $z=z_0$, slightly above the event horizon at $z=0$\cite{Das:2022evy,Das:2023ulz,Das:2023xjr}.\footnote{The proper distance of the Dirichlet brickwall from the event horizon, $\epsilon \approx z_0$ when the wall is close enough to the horizon.} Accordingly, we demand $\phi(z=z_0)=0$, instead of the ingoing boundary condition on the horizon. This boundary condition fixes the ratio of the undetermined constants $C_1$ and $C_2$ of the general solution as
\begin{equation}\label{rat}
    R_{C_2C_1}=\frac{C_2}{C_1}=-z_0^{-2i\alpha} \frac{{}_2F_1(a, b; c; z_0)}{{}_2F_1(1+a-c, 1+b-c; 2-c; z_0)}.
\end{equation}
Following the prescription of \cite{Son:2002sd}, and using the above result, we expand the solution near the boundary at $z = 1$
\begin{equation}
    \phi(z)_{bdry}\sim R_1 (1-z)^{\frac{1}{2}(1-\sqrt{1+\mu^2})} +R_2 (1-z)^{\frac{1}{2}(1+\sqrt{1+\mu^2})},
\end{equation}
where
\begin{align}
    R_1 &=\frac{\Gamma(c)\Gamma(c-a-b)}{\Gamma(c-a)\Gamma(c-b)}+R_{C_2C_1}\frac{\Gamma(2-c)\Gamma(c-a-b)}{\Gamma(1-a)\Gamma(1-b)} \\
    R_2 &= \frac{\Gamma(c)\Gamma(a+b-c)}{\Gamma(a)\Gamma(b)}+R_{C_2C_1} \frac{\Gamma(2-c)\Gamma(a+b-c)}{\Gamma(1+a-c)\Gamma(1+b-c)}
\end{align}
Identifying $R_2$ and $R_1$ as the normalizable and the non-normalizable modes respectively, the boundary Green's function can be computed by taking their ratio
\begin{equation}\label{green}
    G(\omega,m)=\frac{R_2}{R_1}=\frac{ \frac{\Gamma(c)\Gamma(a+b-c)}{\Gamma(a)\Gamma(b)}+R_{C_2C_1} \frac{\Gamma(2-c)\Gamma(a+b-c)}{\Gamma(1+a-c)\Gamma(1+b-c)}}{\frac{\Gamma(c)\Gamma(c-a-b)}{\Gamma(c-a)\Gamma(c-b)}+R_{C_2C_1}\frac{\Gamma(2-c)\Gamma(c-a-b)}{\Gamma(1-a)\Gamma(1-b)}}, 
\end{equation}
with $R_{C_2C_1}$ given in \eqref{rat}.

\begin{figure*}
\begin{subfigure}{0.47\textwidth}
    \centering
    \includegraphics[width=\textwidth]{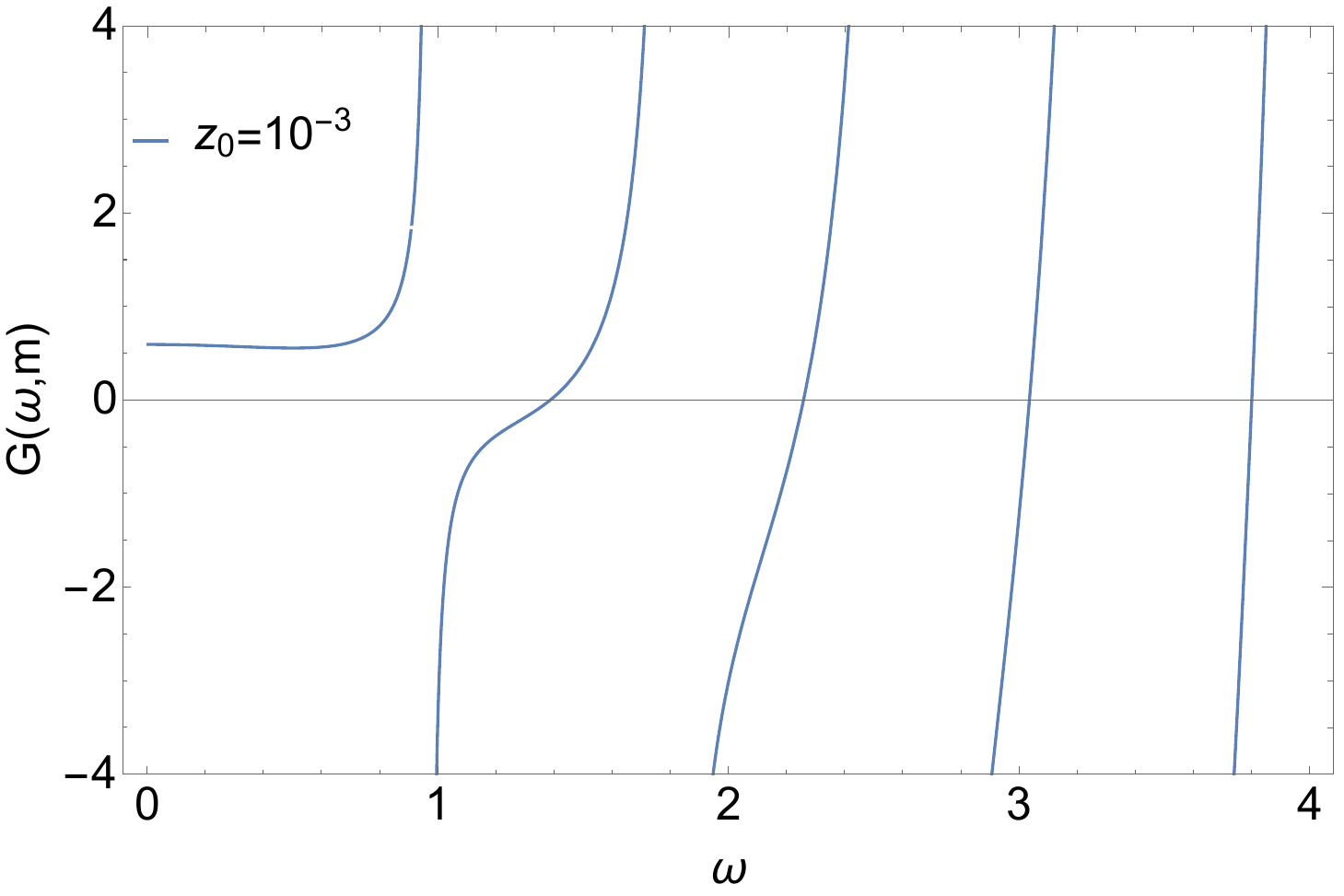}
    \end{subfigure}
    \hfill
    \begin{subfigure}{0.47\textwidth}
    \includegraphics[width=\textwidth]{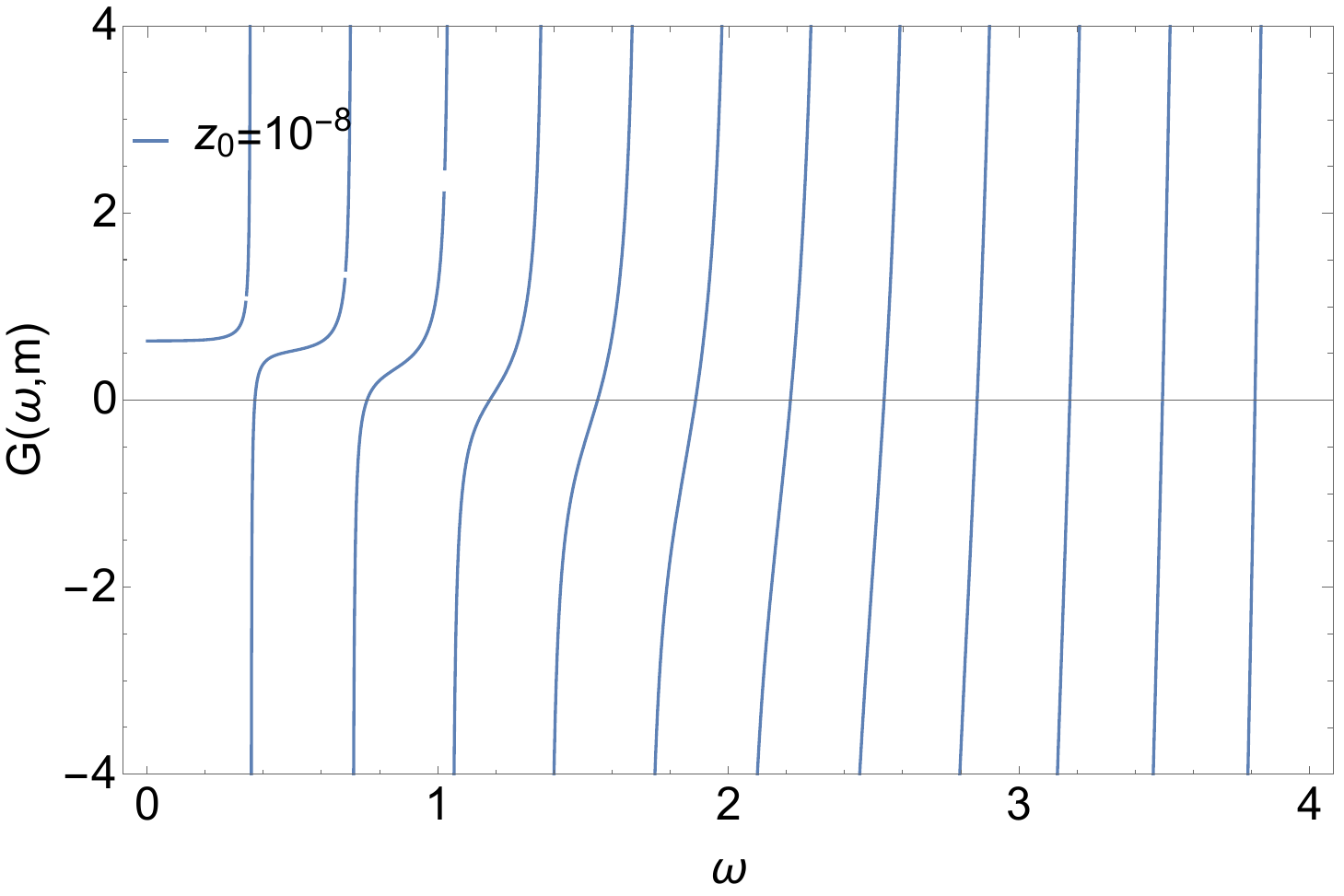}
    \end{subfigure}
    \caption{Pole Structure of $G_{\omega}(n, m)$ for fixed $m=1$. Poles are coming closer and closer as we move the position of the stretched horizon towards the event horizon. $\mu=1$ here for the both figures.}
    \label{green1}
\end{figure*}

The Green's function \eqref{green} possesses a very rich pole structure and the dynamics thereof, as we move the wall close to the horizon of the black hole. As shown in figure \ref{green1}, for a fixed value of $m$, the poles tend to accumulate when the wall is moved closer to the horizon. A very similar dynamics of the poles can also be obtained by keeping $n$ fixed instead of $m$. In what follows, we will investigate this phenomenon of pole accumulation more closely, towards obtaining a physical interpretation of the same.

\subsection{Analytic properties of the Green's function}
The Green's function \eqref{green} has poles when the denominator vanishes\footnote{\eqref{green} also has poles for $a+b-c=-n$ where $n \in \mathcal{Z}_{\geq}0$ which implies $\mu=\sqrt{n^2-1}=0, \sqrt{3}, \sqrt{8}, ...$. We are not interested in those because poles are in the mass axis.}, namely when 
\begin{equation}\label{deno}
    \frac{\Gamma(c)}{\Gamma(c-a)\Gamma(c-b)}+R_{C_2C_1}\frac{\Gamma(2-c)}{\Gamma(1-a)\Gamma(1-b)}=0
\end{equation}
Note, solutions of \eqref{deno} are the normal modes for the probe scalar\cite{Das:2022evy, Das:2023xjr}. This is clearly different from the QNM which is obtained by imposing ingoing boundary conditions. 

When the wall is infinitesimally close to the event horizon, then, from \eqref{rat}, $R_{C_1C_2}\approx -z_0^{-2i\alpha}$. Accordingly, \eqref{deno} can be written as:
\begin{equation}\label{quant2}
    \alpha \log z_0+\text{Arg}\left[\frac{\Gamma(c)}{\Gamma(c-a)\Gamma(c-b)} \right]=n \pi, \hspace{0.1cm}\text{with} \hspace{0.1cm} n\in \mathbf{Z}.
\end{equation}
The solution to \eqref{quant2} provides quantized values of frequency $\omega = \omega_{n,m}$ as the normal mode frequencies. The poles of the Green's function located at these quantized values of frequency get denser as the wall moves closer to the black hole horizon at $z=0$. We already observed this in Figure \ref{green1}, for a fixed value of $m$.

\begin{figure*}
\begin{subfigure}{0.47\textwidth}
    \centering
    \includegraphics[width=\textwidth]{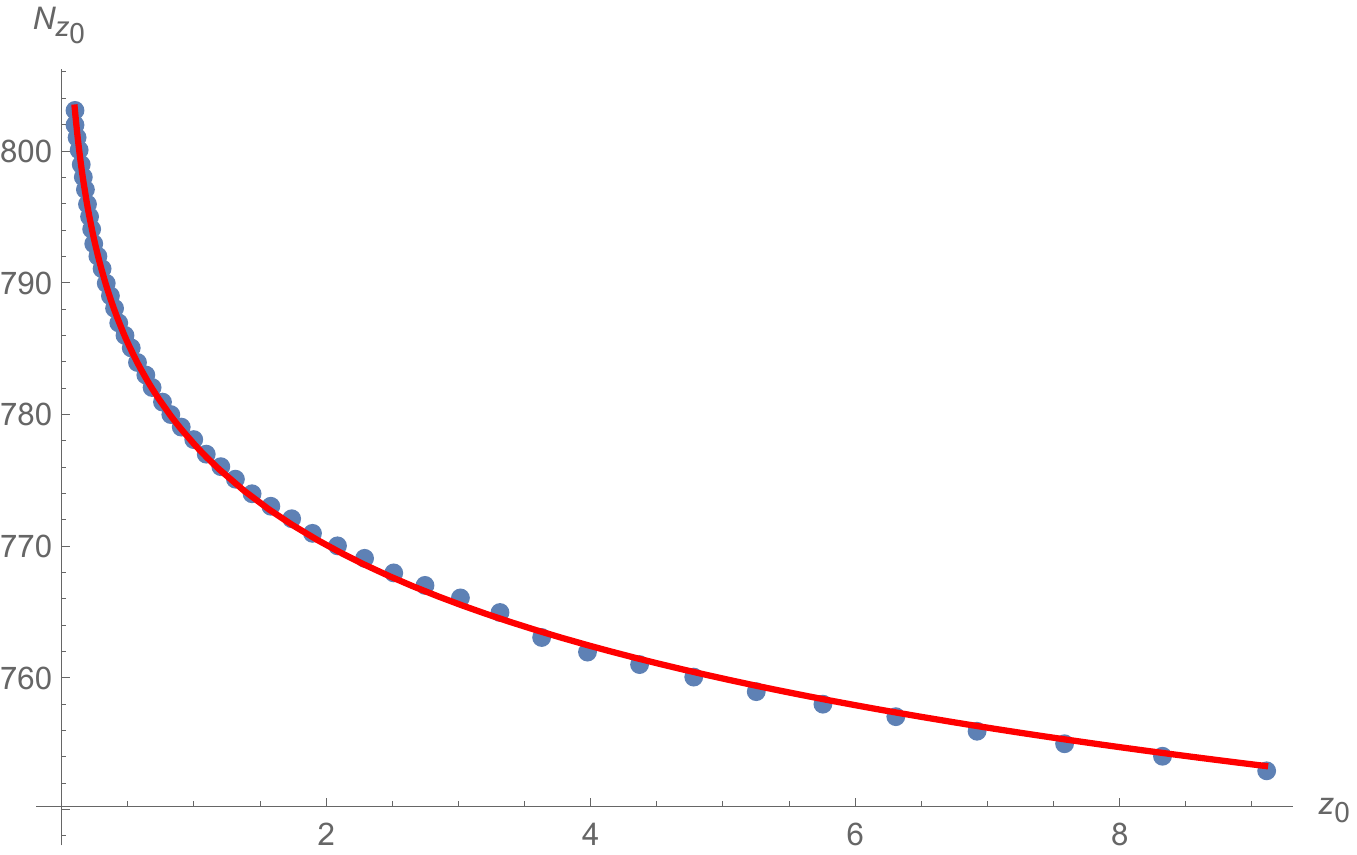}
    \end{subfigure}
    \hfill
    \begin{subfigure}{0.47\textwidth}
    \includegraphics[width=\textwidth]{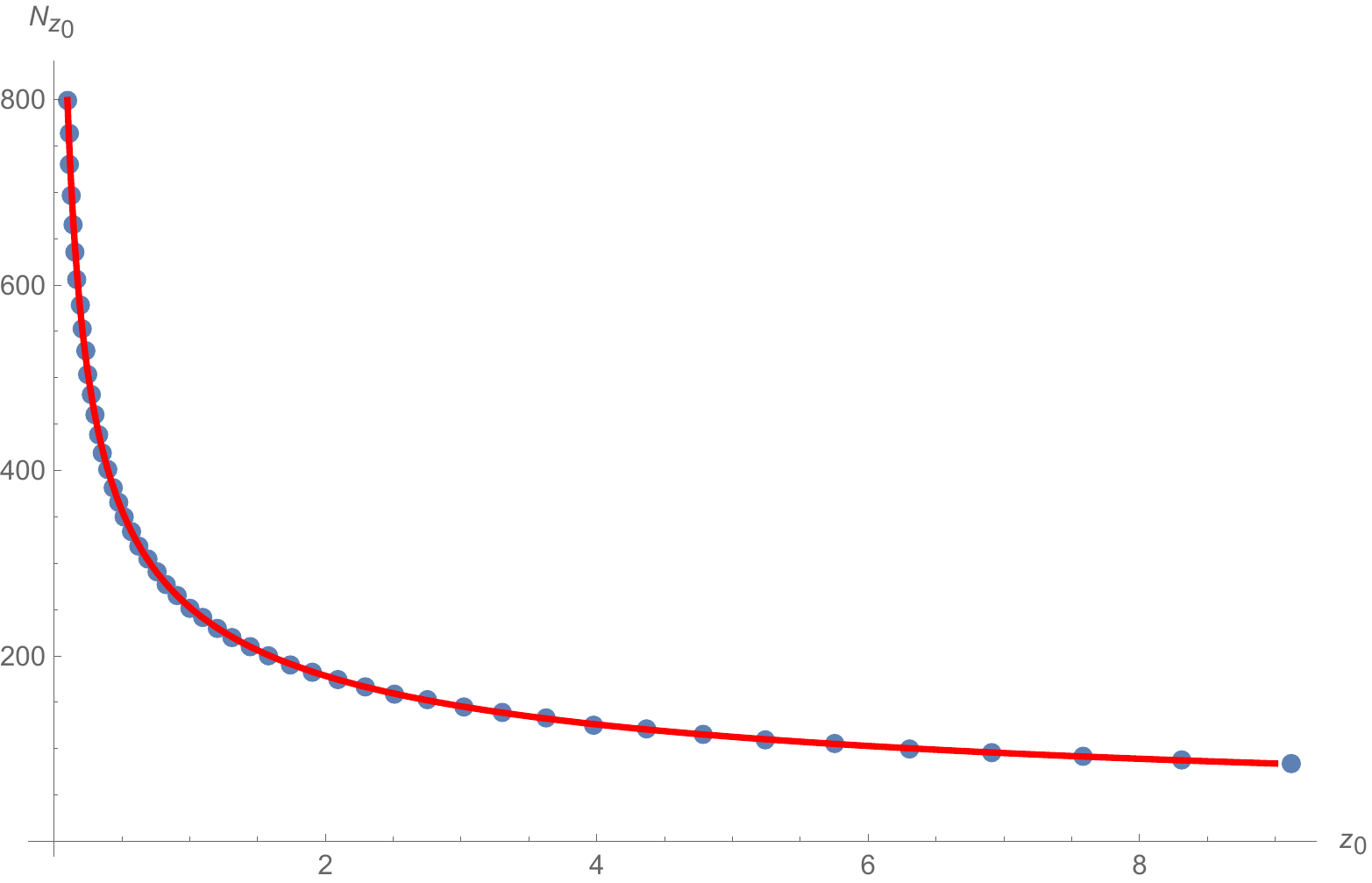}
    \end{subfigure}
    \caption{ $N_{z_0}$ vs. $z_0$ for fixed $m$ (left) and fixed $n$ (right) respectively. $\mu=0$ here. The horizontal axis is scaled up by a factor of $10^{31}$ for better visualization. The numerical data are fitted with appropriate fitting functions which are discussed in the main text.}
    \label{poles1}
\end{figure*}

To get an intuition of how fast that density of poles increases, we can simply fix a cutoff in frequency at $\omega = \omega_{max}$ and count the number of poles as a function of the distance between the wall and the horizon. This estimate can be done both for fixed $m$ and $n$. The results are demonstrated in Figure \ref{poles1}. 
To better visualize, We have fitted the numerical data with appropriate functions. For example, when $m$ is held fixed, the fitting function is
\begin{eqnarray}
a+b \log(1/z_0) \ , \label{ngrow}
\end{eqnarray}
where $\{a,b\}$ are the fitting parameters. This provides the IR-divergent time-scale $\delta=|\log(z_0)|$ where the branch-cut approximation breaks down.  Note that, this time-scale appears on generic kinematic grounds in the presence of a horizon, see {\it e.g.}~\cite{Alam:2012fw} in a different context. On the other hand, when $n$ is fixed, the corresponding fitting function\footnote{The form of this fitting function is inspired from analytic WKB-approximations discussed in \cite{Das:2023xjr}.}
\begin{eqnarray}
a+ b z_0^{-1/2} \ \label{mgrow}
\end{eqnarray}
yields an IR-divergent time-scale, $e^{c \delta}$ with an ${\cal O}(1)$ number $c$.  It is possible to obtain an analytic expression of how dense these poles are along $m$-direction. This result is given by product-log functions and the precise expression involves nested logarithms and powers of them. The existence of the hierarchy is therefore warranted \cite{BDDK}.

It is therefore clear that the rate of accumulation of poles along the $m$-direction is much faster compared to when we fix $m$ to a particular value. Combined with earlier work in \cite{Das:2022evy, Das:2023xjr}, where it was demonstrated how a quantum chaotic feature emerged from level-repulsion along the $m$-direction, the greater accumulation rate along this direction further supports the emerging thermal picture due to the spectral correlations along this direction. It appears to be a rather important qualifier in this framework, where pole accumulation happens along both directions, but level-repulsion occurs in one of them. This will be a very interesting point to understand better. 


\subsection{From discrete poles to a branch-cut: approaching thermality}
\noindent As we discussed in the preceding section, the poles of Green's function accumulate systematically as the Dirichlet wall at $z=z_0$ moves towards the horizon. Approaching $z_0 \rightarrow 0$, the poles form a continuous branch-cut for a low energy asymptotic observer. As we just argued, however, this statement is non-perturbatively enhanced in the presence of non-vanishing angular momentum. We will now demonstrate that this branch-cut approximation encodes the black hole QNM data.
The detailed analyses here are essentially parallel to \cite{Giusto:2023awo}\footnote{Although, the specific system of \cite{Giusto:2023awo} is very different from ours.} and therefore instead of writing them out, we point out the key steps. 

Let us, for simplicity, fix $m$ to a particular value $m=m_0$ for which we denote the normal mode frequencies as $\omega_{n,m_0} \equiv \omega_n$\footnote{As argued before, we can also keep $n$ fixed instead and work with normal mode frequencies $\omega_{n_0,m} \equiv \omega_m$. This will not alter the conclusion.}. In this case, Green's function at fixed $z_0$ can be written in the form of \cite{Giusto:2023awo}
\begin{equation}\label{greenid1}
    G(\omega, m_0)=\sum_{n} \left(\frac{\omega}{\omega_n}  \right)^{\Delta}\frac{\text{Res}(G, \omega_n)}{\omega-\omega_n},
\end{equation}
where ${\text{Res}}(G, \omega_n)$ is the residue of the Green's function at $\omega = \omega_n$.
Here the factor $\omega^\Delta$ is introduced to ensure that the function $\frac{G(\omega,m_0)}{\omega^\Delta}$ falls off and eventually vanishes at the boundary of the contour at $\omega\rightarrow \infty$. When the poles at $\omega = \omega_n$ come infinitesimally close to each other making the structure sufficiently dense, one can instead consider, as a good approximation, a continuum distribution of poles,  simply by replacing the sum over $n$ with an integral over frequency:
\begin{equation}\label{greenid}
    G(\omega, m_0)\approx \int d\omega_n \frac{\rho_{\omega}(\omega_n, m_0)}{\omega-\omega_n},
\end{equation}
where 
\begin{align}
\label{eqn:rho}
 \rho_{\omega}(\omega_n, m_0) =  \left(\frac{\omega}{\omega_n}  \right)^{\Delta}  {\text{Res}}(G, \omega_n)\frac{dn}{d\omega_n}.
\end{align}

From \eqref{greenid}, it is straightforward to show 
\begin{align}
\label{eqn:GR}
    G(\omega+i\epsilon, m_0)- G(\omega-i\epsilon, m_0) &= 2 i\, {\rm{Im}} G_{\cal R}(\omega, m_0) \nonumber \\
    &= 2\pi i \rho_{\omega}(\omega, m_0),
\end{align}
where $G_{\cal R}(\omega, m_0)$ is the retarded Green's function. This shows that, indeed, as the poles become sufficiently dense, the correlator effectively develops a branch-cut with the discontinuity given by the function $\rho_{\omega}(\omega, m_0)$.

The branch-cut discontinuity $\rho_{\omega}(\omega, m_0)$ can be computed from the residue of our Green's function \eqref{green} which yields
\begin{equation}\label{rho}
    \rho_{\omega}(\omega_n, m)=-\frac{1}{\pi}\left(\frac{\omega}{\omega_n}  \right)^{\Delta} \text{Im}\,G_{\text{bh}}(\omega, m)\bigg\rvert_{\omega=\omega_n},
\end{equation}
where $G_{\text{bh}}$ is the retarded Green's function of the BTZ black hole. 
Using \eqref{eqn:GR}, this can be re-expressed as
\begin{align}
\label{eqn:GR-GBH}
2\pi \rho_\omega(\omega, m_0)= {\rm{Im}} \, G_{\cal R}(\omega, m_0) = - \text{Im}\,G_{\text{bh}}(\omega, m_0),
\end{align}
The equation \eqref{eqn:GR-GBH} ensures the matching of the effective temperature in the aforementioned limit to that of a BTZ black hole. This can be confirmed by explicitly calculating the position space representation of Green's function. In particular, one can show {\it in this limit, set by the resolution scale of the low-energy asymptotic observer, the pure state we started with can indeed be well-approximated by a thermal state at the Hawking temperature of the black hole. Accordingly, the position space Green's function becomes indistinguishable from the corresponding thermal correlation function.} It is worth emphasizing here that while our pure state Green's function only contains normal modes, resulting eventually in a branch-cut on the real axis through pole accumulation, \eqref{eqn:GR-GBH} implies that the jump function across the branch-cut actually contains complex QNM poles of the black hole.

A complementary approach to understand this approximate thermality was adopted in  \cite{Burman:2023kko} where the same cutoff scale was obtained by comparing the microcanonical entropy of the horizonless configuration to that associated with a Hartle-Hawking state. This approach is quite similar in spirit to that in \cite{Jafferis:2019wkd}. Connecting our conclusion to this approach, it is worth noting that the entropy of our pure state actually scales with $\sqrt{\delta}$\footnote{see equation 3.10 of \cite{Burman:2023kko}.} so that it represents a system with large entropy in the continuum limit. For such large systems, one can indeed expect a thermal approximation of a typical pure state \cite{lloyd2013pure}.

\section{Algebra and factorization}
The emergence of the approximate thermality in the limit $z_0 \to 0$ carries an underlying algebraic justification.
When the cutoff is placed at a finite $z_0$,
the algebra of observables $\mathcal{A}$ is of type I$_\infty$. Correspondingly, there exists the natural notion of a trace functional as we are used to from quantum mechanics\cite{Witten:2021jzq, Witten:2021unn, Witten:2023qsv}. In the limit $z_0\to 0$, the algebra type transforms to type III \cite{Soni:2023fke}. 
To verify this explicitly, we need to ensure that, in the limit $z_0\to 0$, there is no tracial state $\omega_{\tr}$ on the algebra, i.e.~there is no state satisfying
\begin{align}
    \omega_{\tr}(ab)=\omega_{\tr}(ba)\quad\text{for any}\quad a,b\in\mathcal{A}\,.
\end{align}
 
This can be argued as follows.
Given a finite value of $z_0$, the solution of the scalar wave equation yields a set of quantized simple harmonic oscillator data $\{a_{r,\omega_{n,m}},\omega_{n,m} \}$. Furthermore, we showed that in the limit $z_0\to0$, an effective thermal description emerges at an inverse temperature $\beta=\frac{2\pi l^2}{r_h}$. The associated state vector is given by\cite{Soni:2023fke}
\begin{align}
  \label{eqn:state}  |\beta\rangle_\epsilon=\bigotimes_{\omega_n}\sqrt{1-e^{-\beta\omega_n}}e^{-\beta\frac{\omega_n}{2}a_{l,\omega_n}^\dagger a_{r,\omega_n}^\dagger}|0_{l,\omega_n},0_{r,\omega_n}\rangle\,.
\end{align}
From the perspective of the state, this expression can be thought of as an approximation of the topped-up Boulware vacuum as a Hartle-Hawking state. Such an approximation was argued back in \cite{Mukohyama:1998rf} and was recently revisited through a comparison of the Wightman function in the aforementioned limit in \cite{Burman:2023kko}. This identification is a key to the factorization map discussed in \cite{Jafferis:2019wkd}. The cutoff scale provides an energy topping on the conventional Boulware vacuum, which in the context of dual CFT should have an interpretation of a defect operator. We will elaborate on this interesting connection in a future publication \cite{BDDK}.

Using the vector \eqref{eqn:state}, we evaluate the thermal expectation value of $a_{r,\omega_n}(t)a_{r,\omega_n}^\dagger(0)$ as well as $a_{r,\omega_n}^\dagger(0)a_{r,\omega_n}(t)$ and find that the difference between the two expressions scales as a function of $z_0$.\footnote{The analogous computations may be performed for $a_{l,\omega_n}^{(\dagger)}$.} In particular, we find
\begin{align}
    G(t)=F(t-i \beta)\,,
\end{align}
where,
\begin{align}
    G(t) &= \langle a_{r,\omega_{n,m}}(t)a_{r,\omega_{n,m}}^\dagger(0)\rangle_{\beta,z_0} \nonumber \\
    F(t) &= \langle a_{r,\omega_{n,m}}^\dagger(0)a_{r,\omega_{n,m}}(t)\rangle_{\beta,z_0}\,.
\end{align}
This has the form of the KMS condition, which, in terms of the corresponding Fourier modes $f(\omega)$ and $g(\omega)$ of the functions $F(t)$ and $G(t)$ respectively, translates to
\begin{align}
    g(\omega)=e^{-\beta\omega_{n,m}}f(\omega)\,.
\end{align}
So long as the relative scaling with the exponential is not unity, $F\neq G$ and their difference is non-zero. In this situation one cannot define a trace. In our case, \eqref{ngrow} and \eqref{mgrow} yield
\begin{align}
\label{eqn:approach}
    \left. e^{-\beta\omega_{n,m}}\right|_{m={\rm const}}\sim e^{\frac{\#}{|\ln{z_0}|}} \ , \left. e^{-\beta\omega_{n,m}}\right|_{n={\rm const}}\sim e^{\# z_0^{1/2}} \ ,
\end{align}
where $\#$ denotes an ${\cal O}(1)$ number that does not play any role in the conclusions. 
Now comes the main catch point which will justify the role of the emergent IR time scale mentioned before.
Taking the naive limit $z_0 \to 0$, both the exponential factors tend to 1 and this happens independent of the number $\#$ and for any finite temperature. This implies so long as we can resolve the respective time scales, we can still define a tracial state and accordingly, the algebra remains to be of type I. However, when we are no longer able to do that as the spectrum effectively becomes a continuum with $e^{-\beta\omega_{n,m}} \rightarrow e^{-\beta\omega}$, one cannot define a tracial state for any finite $\beta$.
Accordingly, the algebra type changes to type III.

This emphasizes further the importance of the continuum spectrum approximation in our analysis. Although both the exponential factors in \eqref{eqn:approach} have the same limiting behaviour in the limit $z_0 \to 0$, the gradient of the exponents along $n$ and $m$ directions differ and they determine to what extent this continuous approximation for the spectrum is valid. A faster gradient in the $m$ direction potentially explains the appearance of ramp in the spectral form factor for fixed values of $n$ as observed in \cite{Das:2022evy}.
Thus this serves as an explicit example where the rate of approach of a type I von Neumann algebra to a type III algebra is related to the existence of quantum chaotic behaviour. 

In \cite{Banerjee:2023eew} it was argued that geometric phases associated with the entanglement pattern can be used to distinguish between different algebra types. The trace on any type of algebra is defined by a state vector with vanishing geometric phases, corresponding to maximal entanglement. Therefore, as type III algebras do not have a trace, any state vector for such algebras has geometric phases. It would be interesting to determine whether this notion can be made quantitatively precise in the current setting, i.e.~how the limit $z_0\to0$ affects these geometric phases, in particular their values hinting at an approximate algebraic transition.  A detailed exploration of this will appear in \cite{BDDK}.

\section*{Acknowledgments}
\noindent We thank several discussions with Chethan Krishnan, Samir Mathur, Shiraz Minwalla, Kyriakos Papadodimas, Ronak Soni, Nicholas P.~Warner related to this work. AK is partially supported by CEFIPRA $6304−3$, DAE-BRNS $58/14/12/2021$-BRNS and CRG/$2021/004539$ of Govt.~of India.

\appendix

%

\bibstyle{apsrev4-1}
\bibliography{Bibliography}
\end{document}